\documentclass[fleqn,twoside]{article}
\usepackage[]{espcrc2}
\readRCS
$Id: espcrc2.tex,v 1.2 2004/02/24 11:22:11 spepping Exp $
\usepackage{graphicx}
\usepackage[figuresright]{rotating}

\title{The Six Gluon One-Loop Amplitude}

\author{David C. Dunbar\address[SU]{Department of Physics \\
Swansea University\\
Swansea, SA2 8PP, UK }
 }

\runtitle{One-Loop Six Gluon Amplitude}
\runauthor{David C. Dunbar}

\def\tr{\mathop{\rm tr}\nolimits}

\def\IIs#1#2{{\cal F}
^{{#1}}_{4:#2}}
\def\IIone{\IIs{\rm 1m}}

\def\IIhard{\IIs{{\rm 2m}\,h}}

\def\Fn{4}
\def\Fs#1#2{F^{{#1}}_{\Fn:#2}}
\def\Fone{\Fs{\rm 1m}}
\def\Feasy{\Fs{{\rm 2m}\,e}}
\def\Fhard{\Fs{{\rm 2m}\,h}}

\def\Tr{\mathop{\rm Tr}\nolimits}
\def\L{\left(}
\def\R{\right)}
\def\Wsix#1{W_6^{(#1)}}

\newbox\charbox
\newbox\slabox
\def\s#1{{      
        \setbox\charbox=\hbox{$#1$}
        \setbox\slabox=\hbox{$/$}
        \dimen\charbox=\ht\slabox
        \advance\dimen\charbox by -\dp\slabox
        \advance\dimen\charbox by -\ht\charbox
        \advance\dimen\charbox by \dp\charbox
        \divide\dimen\charbox by 2
        \raise-\dimen\charbox\hbox to \wd\charbox{\hss/\hss}
        \llap{$#1$}
}}

\def\spa#1.#2{\left\langle#1\,#2\right\rangle}
\def\spb#1.#2{\left[#1\,#2\right]}
\def\lor#1.#2{\left(#1\,#2\right)}

\def\cg{c_\Gamma}
\def\rg{r_\Gamma}

\catcode`@=11  

\def\Slash#1{\hskip 0.05 cm \slash\hskip -0.22 cm #1}

\def\Tr{\, {\rm Tr}}

\def\eps{\epsilon}

\def\pol{\eps}

\def\e{\epsilon}

\def\la{\langle}
\def\ra{\rangle}
\def\oneloop{{1 \mbox{-} \rm loop}}

\def\lsl{\not{\hbox{\kern-2.3pt $\ell$}}}
\def\ksl{\not{\hbox{\kern-2.3pt $k$}}}

\def\rg{r_{\Gamma}}

\def\spa#1.#2{\left\langle#1\,#2\right\rangle}
\def\spb#1.#2{\left[#1\,#2\right]}
\def\lor#1.#2{\left(#1\,#2\right)}

\def\sand#1.#2.#3{%
  \left\langle\smash{#1}{\vphantom1}\right|{#2}%
  \left|\smash{#3}{\vphantom1}\right\rangle}
\def\sandp#1.#2.#3{%
  \left\langle\smash{#1}{\vphantom1}^{-}\right|{#2}%
  \left|\smash{#3}{\vphantom1}^{+}\right\rangle}
\def\sandpp#1.#2.#3{%
  \left\langle\smash{#1}{\vphantom1}^{+}\right|{#2}%
  \left|\smash{#3}{\vphantom1}^{+}\right\rangle}
\def\sandmm#1.#2.#3{%
  \left\langle\smash{#1}{\vphantom1}^{-}\right|{#2}%
  \left|\smash{#3}{\vphantom1}^{-}\right\rangle}
\def\sandpm#1.#2.#3{%
  \left\langle\smash{#1}{\vphantom1}^{+}\right|{#2}%
  \left|\smash{#3}{\vphantom1}^{-}\right\rangle}
\def\sandmp#1.#2.#3{%
  \left\langle\smash{#1}{\vphantom1}^{-}\right|{#2}%
  \left|\smash{#3}{\vphantom1}^{+}\right\rangle}

\def\Atree{A^{\rm tree}}
\def\Aloop{A^{\rm 1-loop}}

\def\Lz{\mathop{\hbox{\rm L}}\nolimits_0}
\def\Lt{\mathop{\hbox{\rm L}}\nolimits_2}
\def\Lone{\mathop{\hbox{\rm L}}\nolimits_1}
\def\Kz{\mathop{\hbox{\rm K}}\nolimits_0}

\def\tr{\mathop{\hbox{\rm tr}}\nolimits}

\def\L{\left(}\def\R{\right)}
\def\RP{\right.}
\def\LB{\left[}\def\RB{\right]}

\def\tn#1#2{t^{[#1]}_{#2}}
\def\L{\left(}\def\R{\right)}

\def\BR#1#2{ [#1|{K}|#2\ra}

\def\Gr{{\rm Gr}}

\def\NeqFour{{\cal N} = 4}
\def\NeqOne{{\cal N} = 1}

\newskip\humongous \humongous=0pt plus 1000pt minus 100pt
\def\caja{\mathsurround=0pt}
\def\eqalign#1{\,\vcenter{\openup1\jot \caja
       \ialign{\strut \hfil$\displaystyle{##}$&$\displaystyle{{}##}$\hfil\crcr#1\crcr}}\,} 
\newif\ifdtup

\newcounter{eqnumber}[section]
\renewcommand{\theeqnumber}{\thesection.\arabic{eqnumber}}
\def\equn{\refstepcounter{eqnumber}
\eqno({\rm \theeqnumber})
}

\begin{document}

\begin{abstract}
This article brings together in a single place the different components of the six gluon one-loop amplitude
\vspace{1pc}
\end{abstract}

\maketitle

\section{Introduction}

The one-loop scattering amplitude for six-gluons in pure QCD has been an interesting test case 
for the development of
analytic techniques in perturbative gauge theories. The amplitude can be decomposed into various subamplitudes which were calculated between 1993 and 
2006~\cite{Mahlon:1993si,Bern:1993qk,BDDKa,Bern:1994cg,Bidder:2004tx,Bedford:2004nh,Britto:2005ha,Bern:2005cq,Bern:2005hh,Britto:2006sj,Berger:2006ci,Berger:2006vq,Xiao:2006vt} in thirteen different publications by 23 authors. 

The results for the amplitude are spread across a large number of papers. 
The purpose of this contribution is to bring as many of these as practical together in a common source 
with a reasonably consistent notation. 
There is no original material in this process although some of the original results have been expressed in alternate forms. 

The different pieces have been used in the development of 
various techniques: this article will focus upon the results not the processes by which they were created. 
The amplitudes discussed here are available at $http://pyweb.swan.ac.uk/\sim dunbar/sixgluon.html$ in 
{\tt Mathematica} format.  Many of the helicity amplitudes are included in expressions which are valid for 
$n$ legs.  We will try to show how these specialise to the six gluon case and give the six-gluon case explicitly.

\section{Organisation}

The organisation of loop amplitudes into physical sub-amplitudes is an important step toward computing these amplitudes:
although eventually all the pieces must be reassembled. 
 
For one-loop amplitudes of adjoint representation particles in the
loop, one may perform a colour decomposition similar to the tree-level
decomposition~\cite{ColorDecomposition}.  The one-loop decomposition
is~\cite{Colour},
$$
\eqalign{
{\cal A}_n^{\oneloop}&   = 
\cr
&g^n \sum_{c=1}^{\lfloor{n/2}\rfloor+1}
      \sum_{\sigma \in S_n/S_{n;c}}
     \Gr_{n;c}\L \sigma \R\,A_{n;c}^{}(\sigma)
\cr}
$$
where ${\lfloor{x}\rfloor}$ is the largest integer less than or equal to $x$\,.
The leading colour-structure factor,
$$
\Gr_{n;1}(1) \; = \; N_c\ \Tr\L T^{a_1}\cdots T^{a_n}\R \,,
$$
is just $N_c$ times the tree colour factor, and the subleading colour
structures ($c>1)$ are given by,
$$
\Gr_{n;c}(1) \; = \; \Tr\L T^{a_1}\cdots T^{a_{c-1}}\R\,
\Tr\L T^{a_c}\cdots T^{a_n}\R \,.
$$
$S_n$ is the set of all permutations of $n$ objects
and $S_{n;c}$ is the subset leaving $\Gr_{n;c}$ invariant~\cite{Colour}.
The contributions with fundamental representation quarks can be
obtained from the same partial amplitudes, except that sum runs only
over the $A_{n;1}$ and the overall factor of $N_c$ in $\Gr_{n;1}$ is
dropped.
For one-loop amplitudes of gluons the $A_{n;c}$, $c>1$ can be obtained from the $A_{n;1}$ by summing over 
permutations~\cite{Colour,BDDKa}. Hence it is suffice to compute $A_{n;1}$ in what follows.  
The partial amplitudes $A_{n;1}$
have cyclic symmetry rather than full crossing symmetry

The amplitudes are also organised according to the helicity of the outgoing gluon which may be $\pm$. We
use polarisation tensors formed from Weyl spinors~\cite{Xu:1986xb}
$$\pol^{+}_\mu (k;q) =
{\sandmm{q}.{\gamma_\mu}.k
\over  \sqrt2 \spa{q}.k}\,, \hskip 0.1 cm  
\pol^{-}_\mu (k;q) =
{\sandpp{q}.{\gamma_\mu}.k
\over \sqrt{2} \spb{k}.q}  
$$
where $k$ is the gluon momentum and $q$ is an arbitrary null
`reference momentum' which drops out of final gauge-invariant
amplitudes.  The plus and minus labels on the polarization vectors
refer to the gluon helicities and we use the notation
$\langle ij \rangle\equiv  \langle k_i^{-} \vert k_j^{+} \rangle\, ,
[ij] \equiv \langle k_i^{+} \vert k_j^{-} \rangle$. In the twistor-inspired studies of gauge theory amplitudes
the two component Weyl spinors are often expressed as
$$
\lambda_a =  |k^+\ra \;\;\;\; \bar\lambda_{\dot a} = |k^-\ra
$$ 
Helicity amplitudes are related to those with all legs of opposite helicity by conjugation 
$\spa{a}.b \leftrightarrow \spb{b}.a$. Consequently,
up to conjugation and relabeling,  there are eight independent helicity color ordered amplitudes for the six-gluon amplitude as
given in table~1. Of these amplitudes the $A(1^+2^+3^+4^+5^+6^+)$ and $A(1^-2^+3^+4^+5^+6^+)$ have the simplest one-loop
structure: a consequence that the tree partial amplitudes vanish. In fact, these amplitudes vanish to all orders 
in perturbation theory within any supersymmetry theory.  Amplitudes with exactly two negative helicities are referred to as MHV (``maximally helicity violating'') amplitudes and those with three negative helicities as NMHV 
(``next to MHV'') amplitudes.

Using spinor helicity
leads to amplitudes which are functions of the spinor variables $\spa{a}.b$ and
$\spb{a}.{b}$. It is also useful to define combinations of spinor products
$$
[a|K_{b\cdots f}|m\ra \equiv
\spb{a}.b\spa{b}.m +\cdots +
\spb{a}.f\spa{f}.m 
$$
etc. In terms of Dirac traces
$$
\tr_+ (
\Slash{k_a} \Slash{k_b} \Slash{k_c} \Slash{k_d}
)= \spb{a}.b \spa{b}.c \spb{c}.d \spa{d}.a
$$

 \begin{figure}[t]
\includegraphics*[scale=0.45]{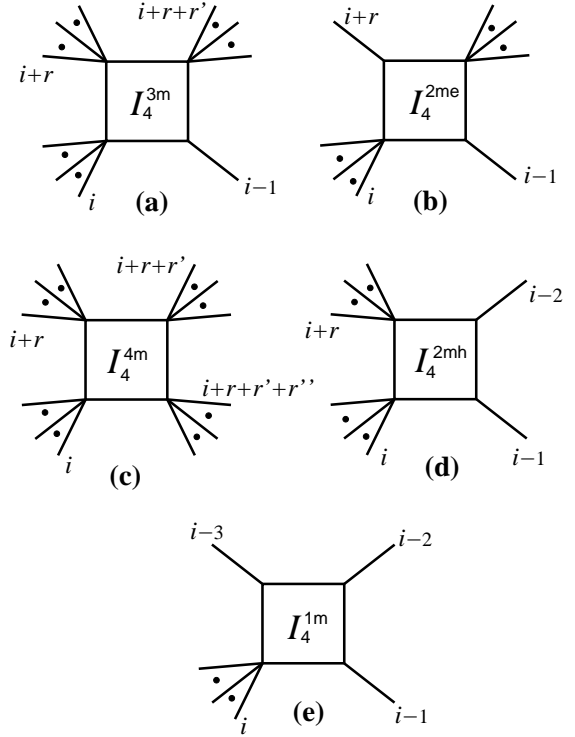}
\caption{The various box integral functions}
	\label{fig:tri}
\end{figure}

A general one-loop amplitude for massless particles can be expressed, after an appropriate Passarino-Veltman reduction~\cite{Passarino:1978jh}, in terms of scalar integral functions with
rational coefficients, 
$$
 \Aloop_n=\sum_{i\in \cal C}\, a_i\, I_4^{i}
 +\sum_{j\in \cal D}\, b_{j}\, I_3^{j}
 +\sum_{k\in \cal E}\, c_{k} \,   I_2^{k}
+R_n\,  ,
\equn\label{DecompBasis}
$$ 
where $a_i,b_i,c_i$ and $R_n$ are rational functions of the $|k_i^{\pm}\ra$ (or equivalently of
$\lambda_a$ and $\bar\lambda_{\dot a}$).  The $I_4$,
$I_3$, and $I_2$ are scalar box, triangle and bubble functions
respectively and these contain the logarithms and dilogarithms of the amplitude. The functional form of the
scalar integrals depends upon the number of legs with non-null momenta inflowing. These are frequently referred to as massive legs although strictly we are dealing with massless states throughout.

We also choose to use a {\it supersymmetric decomposition}. Instead of calculating the one-loop
contributions from massless gluons, $A_n^{[1]}$ or quarks $A_n^{1/2]}$ circulating in the loop it is 
considerable more convenient to calculate the
contributions from a full $N=4$ multiplet, a $N=1$ chiral multiplet and a complex scalar. In terms of these
$$
\eqalign{
A_{n}^{[1]} &\; = \; A_{n}^{\,\NeqFour}-4A_{n}^{\,\NeqOne\; {\rm chiral}}\;+\;A_{n}^{[0]}\,,
\cr
A_{n}^{[1/2]} &\; = \; A_{n}^{\,\NeqOne\; {\rm chiral}}\;-\;A_{n}^{[0]}\,.
\cr}
\equn\label{SusyQCDDecomp}
$$

Finally, we find it 
useful to split an amplitude into its {\it cut-constructible part} and {\it rational part}. The $I_n$ span the
cut-constructible part and
$$
A_n =  C_n  +R_n
$$ This is a convenient rather than a physical split.  The split into
$R_n$ and $C_n$ is, to some extent arbitrary. Some of the integral
functions eg.  the bubble function $I_2$ contain rational pieces
however it is much more logical to keep these within $C_n$.  When
doing so, for the supersymmetric contributions $R_n=0$.  In fact, it
is useful to redefine different integral functions which include as
much of of the rational terms within $C_n$ as possible. When this
happens we will use the notation $\hat R_n$.

Finally, we present out results in the ``four-dimensional-helicity'' (FDH) scheme of dimensional regularisation which 
since it respects supersymmetry merges well with the supersymmetric decomposition.  The translation to
`t Hooft-Veltman scheme is immediate with
$$
A^{N=4\ `tHV}_n =A^{N=4\ FDH}_n-
{ \cg  \over 3} A^{\rm tree }_n 
$$
 
\begin{table}[h]
\hrule
\def\tend{\cr \noalign{ \hrule}}
\halign{
       &  \vrule
       #   \strut        &\strut\hfil #\hfil\vrule
 \strut        &\strut\hfil #\hfil\vrule
 \strut        &\strut\hfil #\hfil\vrule
 \strut         &\strut\hfil #\hfil\vrule
 \strut         &\strut\hfil #\hfil\vrule
       \cr
height15pt  &  Amplitude  & $\;\;\NeqFour  \;\;$        & $\;\; \NeqOne \;\;$   & 
$\;\;\; [0]^C \;\;\;$ &  $\;\;\; [0]^R\;\;\; $ 
\tend
height15pt  & $\; (++++++) \;$  & $0$       &
$0$   &  $0$  & 93~\cite{Bern:1993qk}
\tend
height15pt  & $\; (-+++++) \;$  &    $0$     &
$0$   & $0$  & 93~\cite{Mahlon:1993si}
\tend
height15pt  & $\; (--++++) \;$  & 94~\cite{BDDKa}        &
94~\cite{Bern:1994cg} & 94~\cite{Bern:1994cg} & 05~\cite{Bern:2005cq}
\tend
height15pt  &$\; (-+-+++) \;$  & 94~\cite{BDDKa}        &
94~\cite{Bern:1994cg}  & 04~\cite{Bedford:2004nh} & 06~\cite{Berger:2006vq}
\tend
height15pt  &$\; (-++-++) \;$  & 94~\cite{BDDKa}        &
94~\cite{Bern:1994cg}  & 04~\cite{Bedford:2004nh} & 06~\cite{Berger:2006vq}
\tend
height15pt  &$\; (---+++) \;$  & 94~\cite{Bern:1994cg}        &
04~\cite{Bidder:2004tx}  & 05~\cite{Bern:2005hh} & 06~\cite{Berger:2006ci}
\tend
height15pt  &$\; (--+-++) \;$  & 94~\cite{Bern:1994cg}        &
05~\cite{Britto:2005ha}  & 06~\cite{Britto:2006sj} & 06~\cite{Xiao:2006vt}
\tend
height15pt  &$\; (-+-+-+) \;$  & 94~\cite{Bern:1994cg}        &
05~\cite{Britto:2005ha}  & 06~\cite{Britto:2006sj} & 06~\cite{Xiao:2006vt}
\tend
}
\caption[]{The components of the six-gluon amplitude with the year of computation and original references}
\end{table}

\section{$\NeqFour$ amplitudes}

The one-loop amplitudes for gluon scattering within $\NeqFour$ super-Yang-Mills are particularly simple being heavily constrained by the large symmetry. In terms of the integral basis they can be expressed entirely in terms of scalar boxes~\cite{BDDKa}
$$
 A_n^{\NeqFour}=\sum_{i\in \cal C}\, a_i\, I_4^{i}
$$
The scalar box functions are illustrated in figure~1. 
It is convenient to define rescaled box functions $F_4^{i}$ where
$$
\eqalign{
  I_{4:i}^{1{\rm m}} 
&=
\ -2 \rg {\Fone{i} \over \tn{2}{i-3} \tn{2}{i-2} }
\cr
 I_{4:r;i}^{2{\rm m}e}
&=
\ -2 \rg {\Feasy{r;i}
      \over \tn{r+1}{i-1}\tn{r+1}{i} -\tn{r}{i}\tn{n-r-2}{i+r+1} }\,, 
\cr
I_{4:r;i}^{2{\rm m}h}
&=\ -2 \rg {\Fhard{r;i} \over \tn{2}{i-2} \tn{r+1}{i-1} } \,, 
\cr}
$$
where $\rg=\Gamma(1+\eps)\Gamma^2(1-\eps)/\Gamma(1-2\eps)$. 
The $F$-functions  are the box functions with the appropriate momentum prefactor removed.  For example, 
the scalar box function with one-massive leg is
$$
\eqalign{
F^{1m}_4 &(s_{12},s_{23},m_4^2) 
=
\cr
&
-{1 \over \eps^2} \left( { \mu^2 \over -s_{12}}\right)^{\eps} 
-{1 \over \eps^2}\left( { \mu^2 \over -s_{23}}\right)^{\eps}
+{1 \over \eps^2}\left( { \mu^2 \over -m_4^2 }\right)^{\eps}  
\cr
+ &{\rm Li}_2[ 1 - { -m_4^2 \over -s_{12}}]
+{\rm Li}_2[ 1 - { -m_4^2 \over -s_{23}}]
+{1 \over 2} \ln^2( {-s_{12}\over -s_{23} } )
\cr}
$$
These integrals are expressed in the Euclidean region where all momentum invariants are negative i.e.
$-s > 0$ etc. $m_4^2$ is the momentum invariant inflowing to the fourth leg not a physical mass.    

The expressions are given in the physical region through the usual analytic continuation. 
We use notations,
$$
t_j^{[m]} = K_{j\cdots j+m}^2 \equiv (k_j+k_{j+1}+\cdots +k_{j+m} )^2
$$
but also the shorthand $s_{ij}=(k_i+k_j)^2$ and $t_{ijk}=(k_i+k_j+k_k)^2$.

Closed analytic expressions are known for the $\NeqFour$ MHV and NMHV $n$-point one-loop amplitudes. The MHV amplitudes are given by
$$
A_n^{\NeqFour,MHV}=
\cg \Atree_n \times \biggl(  \sum_i F_4^{1m}   
+\sum_i  F_4^{2me}  
\biggr)
$$
where the sum over $F$-functions is over all possible inequivalent 
functions with the appropriate cyclic ordering of legs.  
For the six-point amplitude this consists of the six one-mass box functions and the three independent
$F_4^{2m e}$ box functions. 
The three different MHV amplitudes only differ by the overall $\Atree_6$ factor.  
The amplitude has an overall factor of
$$
c_\Gamma =  { (4\pi)^{\eps} \over 16\pi^2 } {
\Gamma ( 1+\eps) \Gamma^2(1-\eps) \over \Gamma (1-2\eps) }
={ \rg \over (2\pi)^{2-\eps} } 
$$
In subsequent amplitudes we will usually suppress this factor.   
In terms of elementary functions we can express the six-point amplitude as
$$
\eqalign{
&\biggl(  \sum_i F_4^{1m}   
+\sum_i  F_4^{2me}  
\biggr)
=\cr
&\sum_{i=1}^{6}  -{ 1 \over \eps^2 } \biggl(
{ \mu^2  \over -\tn2{i} } \biggr)^{\eps}
-\sum_{i=1}^6
  \ln \biggl({ -\tn{2}{i}\over -\tn{3}{i} }\biggr)
  \ln \biggl({ -\tn{2}{i+1}\over -\tn{3}{i} }\biggr) 
\cr
& 
-{\rm Li}_2 \biggl[1-{ \tn{2}{1} \tn{4}{6}\over
\tn{3}{1} \tn{3}{6} } \biggr]
-{\rm Li}_2 \biggl[1-{ \tn{2}{2} \tn{4}{1}\over
\tn{3}{2} \tn{3}{1} } \biggr]
\cr
&-{\rm Li}_2 \biggl[1-{ \tn{2}{3} \tn{4}{2}\over
\tn{3}{3} \tn{3}{2} } \biggr]\  +
\cr
&-{1\over 4} \sum_{i=1}^6
  \ln \biggl({ -\tn{3}{i}\over -\tn{3}{i+4}  } \biggr)
  \ln \biggl({ -\tn{3}{i+1}\over -\tn{3}{i+3} } \biggr)\ +{  \pi^2}\
\cr}
$$

The three independent NMHV amplitudes also contain the one-mass box functions but contain $F_4^{2m h}$ boxes. 
The boxes which appear in the six-point NMHV amplitudes appear in the very special combination
$$
\eqalign{
  & \Wsix{i}\ \equiv\ \Fone{i} + \Fone{i+3}
                 + \Fhard{2;i+1} + \Fhard{2;i+4} \cr
     \ &=\ -{1\over2\e^2} \sum_{j=1}^6
         \left( { \mu^2 \over -s_{j,j+1} } \right)^\e
 \cr
&  -\ \ln\left({-t_{i,i+1,i+2} \over -s_{i,i+1}}\right)
      \ln\left({-t_{i,i+1,i+2} \over -s_{i+1,i+2}}\right)\cr
 &\quad
  \ -\ \ln\left({-t_{i,i+1,i+2} \over -s_{i+3,i+4}}\right)
      \ln\left({-t_{i,i+1,i+2} \over -s_{i+4,i+5}}\right)
 \cr
& \ +\ \ln\left({-t_{i,i+1,i+2} \over -s_{i+2,i+3}}\right)
      \ln\left({-t_{i,i+1,i+2} \over -s_{i+5,i}}\right)\cr
 &\quad
 \ +\ {1\over 2}\ln\left({-s_{i,i+1} \over -s_{i+3,i+4}}\right)
         \ln\left({-s_{i+1,i+2} \over -s_{i+4,i+5}}\right)
\cr
&\ +\ {1\over 2}\ln\left({-s_{i-1,i} \over -s_{i,i+1}}\right)
         \ln\left({-s_{i+1,i+2} \over -s_{i+2,i+3}}\right)\cr
  &\quad
 \ +\ {1\over 2}\ln\left({-s_{i+2,i+3} \over -s_{i+3,i+4}}\right)
         \ln\left({-s_{i+4,i+5} \over -s_{i+5,i}}\right)
 \ +\ {\pi^2\over3}\ . \cr}
\equn\label{Wdef}
$$
As we can see the dilogarithms drop out of this expression. It is an example where the expansion in terms 
of scalar boxes is probably not optimal in some sense. This feature of the $\NeqFour$ NMHV six-point 
amplitude persists 
for amplitudes involving external states other than gluons~\cite{Bidder:2005in} but not beyond six-points. 

The first NMHV $\NeqFour$ amplitude is given by
$$
\eqalign{
A_{6;1}^{N=4}(& 1^-2^-3^-4^+5^+6^+)\ 
\cr
&=i\cg\ \LB B_1\,\Wsix1+B_2\,\Wsix2+B_3\,\Wsix3\RB
\cr}
\equn\label{pppmmmloop}
$$
The coefficients $B_i$ are given in terms of the $B_0$ function
$$
\eqalign{
B_0\ &=\  {
  \,  t_{123}^3
  \over \spb1.2\spb2.3\spa4.5\spa5.6\
    [1|K_{23}|4\ra [3|K_{12}|6\ra }
 \ . \cr}
$$
by
$$
\eqalign{
  B_1\ &=\ B_0 \ , \cr
  B_2\ &=\ \left({ [4|K_{123}|1\ra 
         \over t_{234} } \right)^4  \ B_0^+
       + \left({ \spa2.3\spb5.6 \over t_{234} } \right)^4
            \ \bar{B}_0^+
          \ , \cr
  B_3\ &=\ \left({ [6|K_{345}|3\ra 
         \over t_{345} } \right)^4  \ B_0^- 
       + \left({ \spa1.2\spb4.5 \over t_{345} } \right)^4
            \ \bar{B}_0^-  \ . \cr}
$$
where
$$
B_0^+ \equiv B_0|_{i \longrightarrow i+1} \;\;
B_0^- \equiv B_0|_{i \longrightarrow i-1} 
$$
and 
$$
\bar{B}_0
\equiv
B_0 | _{ \spa{a}.b \leftrightarrow \spb{a}.b }
$$
i.e. conjugation.

The other amplitudes are
$$
\eqalign{
A_{6;1}^{N=4}& (1^-2^- 3^+4^-5^+6^+)\ 
\cr
&=\
i \cg\ \left[ D_1\,\Wsix1 + D_2\,\Wsix2 + D_3\,\Wsix3 \right],
\cr}
$$
where
$$
\eqalign{
  D_1\ &=\ \left({ [3|K_{123}|4\ra
         \over t_{123} } \right)^4    \ B_0
       + \left({ \spa1.2\spa5.6 \over t_{123} } \right)^4
            \ \bar{B}_0 \ , \cr
  D_2\ &=\ \left({ [3|K_{234}|1\ra 
         \over t_{234} } \right)^4    \ B_0^+ 
       + \left({ \spa2.4\spb5.6 \over t_{234} } \right)^4
            \ \bar{B}_0^+   \ , \cr
  D_3\ &=\ \left(
{ [6|K_{345}|4\ra 
         \over t_{345} } 
\right)^4    \ B_0^- 
       + \left(
{ \spa1.2\spb3.5 \over t_{345} } \right)^4
             \bar{B}{}_0^-  , 
\cr}
$$
and
$$
\eqalign{
A_{6;1}^{N=4}& (1^-2^+3^-4^+5^-6^+)
\cr
&= i \cg\ \left[ G_1\,\Wsix1 + G_2\,\Wsix2 + G_3\,\Wsix3 \right],
\cr}
$$
where
$$
\eqalign{
  G_1\ &=\ \left({ [2|K_{456} | 5 \ra
         \over t_{123} } \right)^4    \ B_0
       + \left({ \spa1.3\spb4.6 \over t_{123} } \right)^4
            \ \bar{B}_0 \ , \cr
  G_2\ &=\ \left({ [6|K_{234} | 3 \ra 
 \over t_{234} } \right)^4    \bar{B}_0^+ 
       + \left({ \spa5.1\spb2.4 \over t_{234} } \right)^4
            \ B_0^+ \ , \cr
  G_3\ &=\ \left({ [4|K_{612} | 1\ra 
         \over t_{345} } \right)^4    \bar{B}_0^-
       + \left({ \spa3.5\spb6.2 \over t_{345} } \right)^4
            \ B_0^- \ . \cr}
$$

\section{$\NeqOne$ amplitudes $(--++++)$,$(---+++)$}

The simplest of the non-zero $\NeqOne$ amplitudes is the MHV with the
two-minus helicities arranged adjacent to each other~\cite{Bern:1994cg}. For this
configuration the amplitude contains no boxes (or triangles) and is given by
$$
\eqalign{
A^{N=1} & (1^-2^-3^+ \ldots n^+) =
\cr
& \hskip -0.5 truecm
{\cg  A^{\rm tree}  \over2}
\biggl\{
\biggl(
\Kz ( -t_2^{[2]}/\mu^2 ) + \Kz( -t_n^{[2]}\mu^2 )
\biggr)
\cr
 - {1 \over  \tn{2}1 }
&
   \sum_{m=4}^{n-1} c_{12}^m {\Lz \L -\tn{m-2}2/(-\tn{m-1}2)\R\over \tn{m-1}2}
   \biggr\}
\cr}
$$
where
$$
c_{12}^m = \Bigl( \tr_+[\Slash{k_1}\Slash{k_2}\Slash{k_{m}}\Slash{q_{m,1}} ]
-\tr_+[\Slash{k_1}\Slash{k_2}\Slash{q_{m,1}}\Slash{k_{m}} ] \Bigr)
$$
using 
$$
q_{m,l}=\left\{\eqalign{\sum_{i=m}^{l} k_i,&\qquad m \leq l ,\cr
                   \sum_{i=m}^{n} k_i + \sum_{i=1}^l k_i,&\qquad m>l,\cr}\RP
$$
and
$$
\Lz (r) = { \ln (r) \over 1-r } 
\;\;
\Kz (r) =  {1 \over \eps } -\ln(r)  +2
$$

\begin{figure}[t]
\includegraphics*[scale=0.45]{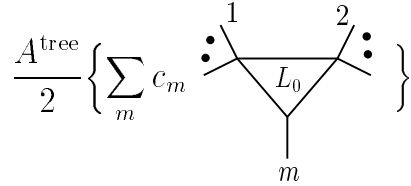}
\caption{The N=1 one-loop contribution to the simplest of the MHV amplitudes
$A(1^-2^-3^+4^+\cdots n^+)$. The $\Lz$ function can be thought of as either a combination of bubble functions or as arising from 
a Feynman parameter integral of the two mass triangle indicated. }
	\label{agjacentMHV}
\end{figure}

The $\Kz$ function is just the bubble integral function $I_2(s)=\Kz(-s/\mu^2)$. 
We have expressed the amplitude using $\Lz(s/s')/s'$ expressions
to avoid spurious singularities. In the limit $s/s' \longrightarrow 1$,
$$
{
\Lz(-s/-s')
\over s' }
={ \ln(-s) -\ln(-s')
\over s'-s  }
\longrightarrow -1
$$
which is obviously non-singular however if we express the amplitude in terms of the scalar bubble
functions $I_2$ using
$$
{
\Lz(-s/-s')
\over s' }
={ I_2(-s') -I_2(-s) 
\over s'-s  }
$$ then the coefficients of $I_2(s)$ and $I_2(s')$ both contain the
spurious singularity. 
For the six-point case there are just two $\Lz$ functions in the summations and
$$
\eqalign{
A^{N=1} & (1^-2^-3^+4^+5^+ 6^+) =
\cr
& \hskip -0.5 truecm
{ \cg A^{\rm tree}  \over2}
\biggl\{
\biggl(
\Kz ( -s_{23}/\mu^2 ) + \Kz( -s_{61}/\mu^2 )
\biggr)
\cr
 & \hskip -1.0 truecm     -{ c_{12}^4 \over  s_{12} }{\Lz \L -s_{23}/(-t_{234})\R\over t_{234}}
 -{c_{12}^5 \over  s_{12} }{\Lz \L -s_{61}/(-t_{234})\R\over t_{234}}
\biggr\}
\cr}
$$

The next simplest amplitude we present is not one of the remaining MHV amplitudes but the amplitude $A(---+++)$. This example of a ``split helicity'' configurations contains many simplifications and can be expressed using  $\Lz$ and $\Kz$ functions. 
The amplitude is symmetrical under the operations 
$$
\eqalign{
A & \longrightarrow A|_{123456\longrightarrow 321654} 
\cr
\;\;A & \longrightarrow \bar{A}|_{123456 \longrightarrow 456123} 
\cr}
$$

 There is an all-$n$ expression for the configuration with three adjacent negative helicities~\cite{Bidder:2005ri}, 
$$
\eqalign{
&A_{n}^{\,\NeqOne\ {\rm chiral}}(1^-2 ^-3^-4^+5^+\cdots n^+)
\; = 
\cr
\; &
{\cg \Atree \over 2} \,\left( \Kz( -s_{n1}/\mu^2 ) +\Kz( -s_{34}/\mu^2 )
\right)
\cr
&
-{i\cg \over 2} \biggl( 
\sum_{r=4}^{n-1} \,\hat d_{n,r}\,
{    \Lz [ q_{3,r}^2 / q_{2,r}^2 ] \over q_{2,r}^2  }
\cr
&\hspace{-0.6cm}
+
\sum_{r=4}^{n-2}\, \hat g_{n,r}\,
{    \Lz [ q_{2,r}^2 / q_{2,r+1}^2 ] \over q_{2,r+1}^2  }
+
\sum_{r=4}^{n-2}\, \hat h_{n,r}\,
{    \Lz [ q_{3,r}^2 / q_{3,r+1}^2 ] \over q_{3,r+1}^2  } \biggr) 
\cr}
$$
where
$$
\eqalign{
\hat d_{n,r}\; = \; &
{  \la 3 | {q}_{3,r}{q}_{2,r}  | 1 \ra^2\,
\la 3 | {q}_{3,r}\big[k_2,{q}_{2,r}\big]{q}_{2,r}  | 1 \ra \spa{r}.{r+1}
\over
 [ 2 | {q}_{2,r}  | r \ra
 [ 2 | {q}_{2,r}  | {r+1} \ra\, 
 \prod_{k=3}^{n} \spa{k}.{k+1}  \,{q}_{2,r}^2\,q_{3,r}^2}\,,
\cr
\hat g_{n,r}\ =\ & 
\cr
& \hskip -1.25 truecm \sum_{j=4}^{r}   
{\la 3|q_{3,j} {q}_{2,j}|1\ra^2  
\la 3|q_{3,j}{q}_{2,j}\big[k_{r+1},{q}_{2,r}\big]|1\ra
\spa{j}.{j+1} 
\over
[2|q_{2,j}|j\ra
[2|q_{2,j}|j+1\ra 
\prod_{k=3}^{n}\spa{k}.{k+1} \,q_{3,j}^2\,q_{2,j}^2} 
 \cr
\hat h_{n,r}\; = \; & (-1)^n \,
\hat g_{n,n-r+2}\bigl\vert_{(123..n)\to(321n..4)}\,.
\cr}
$$

This general expression reduces to the explicit form for the six-point case, 
$$
\eqalign{& A_6^{N=1}(  1^-2^-3^-4^+5^+6^+) = 
\cr
& a_1 \Kz[-s_{61}/\mu^2] +a_2
\Kz[-s_{34}/\mu^2] \cr 
& - {i\cg \over 2} \Biggl[ c_1 { \Lz [t_{345}/ s_{61} ] \over s_{61} } +c_2{
\Lz [ t_{234} /s_{34} ] \over s_{34}}\cr 
& +c_3{ \Lz [ t_{234}/s_{61} ]
\over s_{61}} +c_4{ \Lz [ t_{345} /s_{34} ] \over s_{34} } 
\Biggr] \cr}
$$
where the coefficients are
$$
a_1 =a_2 = {\cg\over2} \Atree_6(1^-2^-3^-4^+5^+6^+),
$$
and
$$
\eqalign{ c_1 =& {  [6|K|3\ra^2
[ 6 | ( k_2 K  -Kk_2) K|3\ra  \over
[2|K|5\ra  \spb6.1\spb1.2\spa3.4\spa4.5 K^2 } \,, 
\;
 K=K_{345} \cr 
\cr
c_2 =& c_1|_{123456\longrightarrow 321654}
\;\;
c_4 = c_3|_{123456\longrightarrow 321654}
\cr
& \hskip 1.0 truecm c_3=  \bar{c}_1|_{123456\longrightarrow 654321}
\cr}
$$

\section{Basis of Box Functions}

At this point we must
discuss a suitable basis for expressing the amplitudes. We could use the basis
(\ref{DecompBasis}) however this is not the most efficient option. 
By choosing a suitable basis of box functions we can considerably simplify the structure of the triangle coefficients.

Triangle integral functions may have one, two or three massless legs: 
$I_{3}^{2\rm m}$, $I_{3}^{1\rm m}$, $I_{3}^{3\rm m}$. 
The one-mass triangle depends only on the momentum invariant of the
massive leg $K_1$ and is
$$
I_{3}^{1\rm m} = {\rg\over\e^2} (-K_1^2)^{-1-\eps} \ .
$$ whilst 
the two-mass triangle integral with non-null momenta $K_1$ and $K_2$ is,
$$
I_{3}^{2 \rm m} = {\rg\over\e^2}
{(-K_1^2)^{-\eps}-(-K_2^2)^{-\eps} \over  (-K_1^2)-(-K_2^2) }\ .
$$ Both these integral functions contain $\ln(K^2)/\eps$ IR
singularities.  The key point is that the IR singularity of an
amplitude must be~\cite{Kunszt:1994mc}
$$
A^{N=1\ \rm chiral}_{IR} =  {\cg \over \eps} \Atree
\;\;\;\;\;
A^{[0]}_{IR} =  {\cg \over 3 \eps} \Atree
$$ so that the $\ln(K^2)/\eps$ singularities must cancel.  This
constraint effectively determines the coefficients of $I_{3}^{1\rm m}$
and $I_{3}^{2\rm m}$ in terms of the box coefficients.

Specifically, the one and two mass triangles are linear combinations of
the set of functions,
$$
G(-K^2)= \rg { (-K^2)^{-\eps}  \over\e^2} \; ,
$$
with
$$
\eqalign{
I^{1m}_3 &= G(-K_1^2)\;\; ,
\;\;
\cr
I^{2m}_3 &={ 1 \over (-K_1^2)-(-K_2^2) }
\left( G(-K_1^2)-G(-K_2^2  ) \right).
\cr}
$$
The $G(-K^2)$ are labeled by the independent momentum invariants
$K^2$ and in fact form an independent basis of functions, unlike the
one and two-mass triangles which are not all independent.

In practice we need never calculate the coefficients of the $G$ functions once we know the 
box coefficients. 
The only functions containing $\ln(s)/\eps$ terms are the box functions and $I_4^{1m}$ and $I_4^{2m}$ so
$$
\sum a_i I_4|_{\ln(K^2)/\eps}  + b_G  { \ln(K^2) \over \eps }  
 = 
0
$$
This equation fixes the single $b_G$ in terms of the $a_i$.

The simplest approach to implement this simplification is to express
the amplitude in terms of truncated finite $F$-functions.  If we
define the function
$$
\eqalign{
&{\cal F}^{1m}_4 (s_{12},s_{23},K_4^2) = F^{1m}_4 (s_{12},s_{23},K_4^2) 
\cr
&+{1 \over \eps^2} \left( { \mu^2 \over -s_{12}}\right)^{\eps} 
+{1 \over \eps^2}\left( { \mu^2 \over -s_{23}}\right)^{\eps}
-{1 \over \eps^2}\left( { \mu^2 \over -K_4^2 }\right)^{\eps}  
\cr}
$$
and use these, together with the other truncated functions, as a basis then the $\NeqOne$ and scalar amplitudes can be expressed as
$$
\Aloop =  \sum  a_i {\cal F}^i_4 + \sum b_j^{3m} I_3^{3m,j} + \sum c_k I_2^k  +R
$$
with no $I_3^{1m}$ and $I_3^{2m}$  present.
The truncated two-mass-easy box functions is
$$
\eqalign{
{\cal F}^{2me}_4 & (S,T,K_2^2,K_4^2)
= F^{2me}_4 (S,T,K_2^2,K_4^2)
\cr
&+{1 \over \eps^2} \left( { \mu^2 \over -S}\right)^{\eps} 
+{1 \over \eps^2}\left( { \mu^2 \over -T}\right)^{\eps}
\cr
&-{1 \over \eps^2}\left( { \mu^2 \over -K_2^2 }\right)^{\eps}  
-{1 \over \eps^2}\left( { \mu^2 \over -K_4^2 }\right)^{\eps}  
\cr}
$$  
where $S=(k_1+K_2)^2$ and $T=(K_2+k_3)^2$.  
This function was labeled $M_0(S,T,K_2^2,K_4^2)$ in ref.~\cite{Bern:1994cg} and fig.~2. It has the feature that
its soft limit is smooth
$$
{\cal F}^{2me}_4 (S,T,0,K_4^2)=
{\cal F}^{1m}_4 (S,T,K_4^2)
$$
which means the $F^{2me}$ and $F^{1m}$ can be combined in a single summation.  
The 
truncated two-mass-hard box functions is
$$
\eqalign{
{\cal F}^{2mh} & (S,T,K_3^2,K_4^2)
= F^{2mh} (S,T,K_3^2,K_4^2)
\cr
&+{ 1\over 2\eps^2} \left( { \mu^2 \over -S}\right)^{\eps} 
+{1 \over \eps^2}\left( { \mu^2 \over -T}\right)^{\eps}
\cr
&-{1 \over 2\eps^2}\left( { \mu^2 \over -K_3^2 }\right)^{\eps}  
-{1\over 2\eps^2}\left( { \mu^2 \over -K_4^2 }\right)^{\eps}  
\cr}
$$  

\section{$\NeqOne$ amplitudes $(-+-+++)$,$(-++-++)$}

These are the two remaining MHV amplitudes.
$$
A(1^-2^+3^-4^+5^+6^+)\;\;
A(1^-2^+3^+4^-5^+6^+)
$$
These two amplitudes can be expressed in a general way. 
Suppose we have a MHV amplitude with the negative helicity legs being 
$i$ and $j$ then the amplitude is a combination of $F^{2me}_4$, $F^{1m}$, $\Lz$ and $\Kz$ functions
$$
\Atree \times \biggl( 
\sum a_{ij}^{n_1n_2} {\cal F}^{2me}_4 
+\sum c^{ij}_{m,a} \Lz \biggr)
$$
In the above the functions ${\cal F}^{1m}$ and $K_0$ are implicit as special cases. When one of the Kinematic invariants is zero we {\it replace} $\Lz(s/s')/s'$ by $\Kz(-s/\mu^2)/s$. This is {\it not} a smooth limit.  
The boxes which are present are a restricted set of the $I_4^{2m\ e}$ and  $I_4^{1m}$. To be included in the sum
the two negative helicities must lie in the two massive legs of the integral function. 
For such a $I_4^{2m\ e}$ if we label the two massless legs by $n_1$ and $n_2$ 
then the coefficient of this box is 
$$
a_{ij}^{n_1n_2}
=
2{ \spa{n_1}.{i} \spa{n_1}.{j} \spa{n_2}.{i} \spa{n_2}.{j}
\over  \spa{i}.{j}^2 \spa{n_1}.{n_2}^2  }
$$
The formula also holds when one of the two masses is zero. 
For the $\Lz$ functions there is a visual  realisation of the summation if we recognize that $\Lz$ is also the result of 
carrying out a Feynman parameter integral of a two-mass triangle $\Lz(s/s')/s'=I_3^{2m} (s,s')[a_2]$.   In this case
the summation runs over $\Lz$ where  one of the negative helicities lies in each of the massive legs.  
The general summation formula is given in ref.~\cite{Bern:1994cg}. The  coefficient is given by
$$
\eqalign{
 c^{ij}_{m,a}
&= {(
{\tr_{+}} [ \Slash{k_i}\Slash{k_j} \Slash{k_m}\Slash{q_{m,a}}]
-{\tr_{+}} [ \Slash{k_i}\Slash{k_j} \Slash{q_{m,a}}\Slash{k_m}]
) \over \LB (k_i + k_j )^2\RB^2 }
\cr
&\times 
\spa{m}.i\spb{i}.j\spa{j}.m\,
{ \spa{a}.{a+1} \over \spa{a}.m \spa{m}.{a+1} }
}
$$

\begin{figure}[t]
\includegraphics*[scale=0.4]{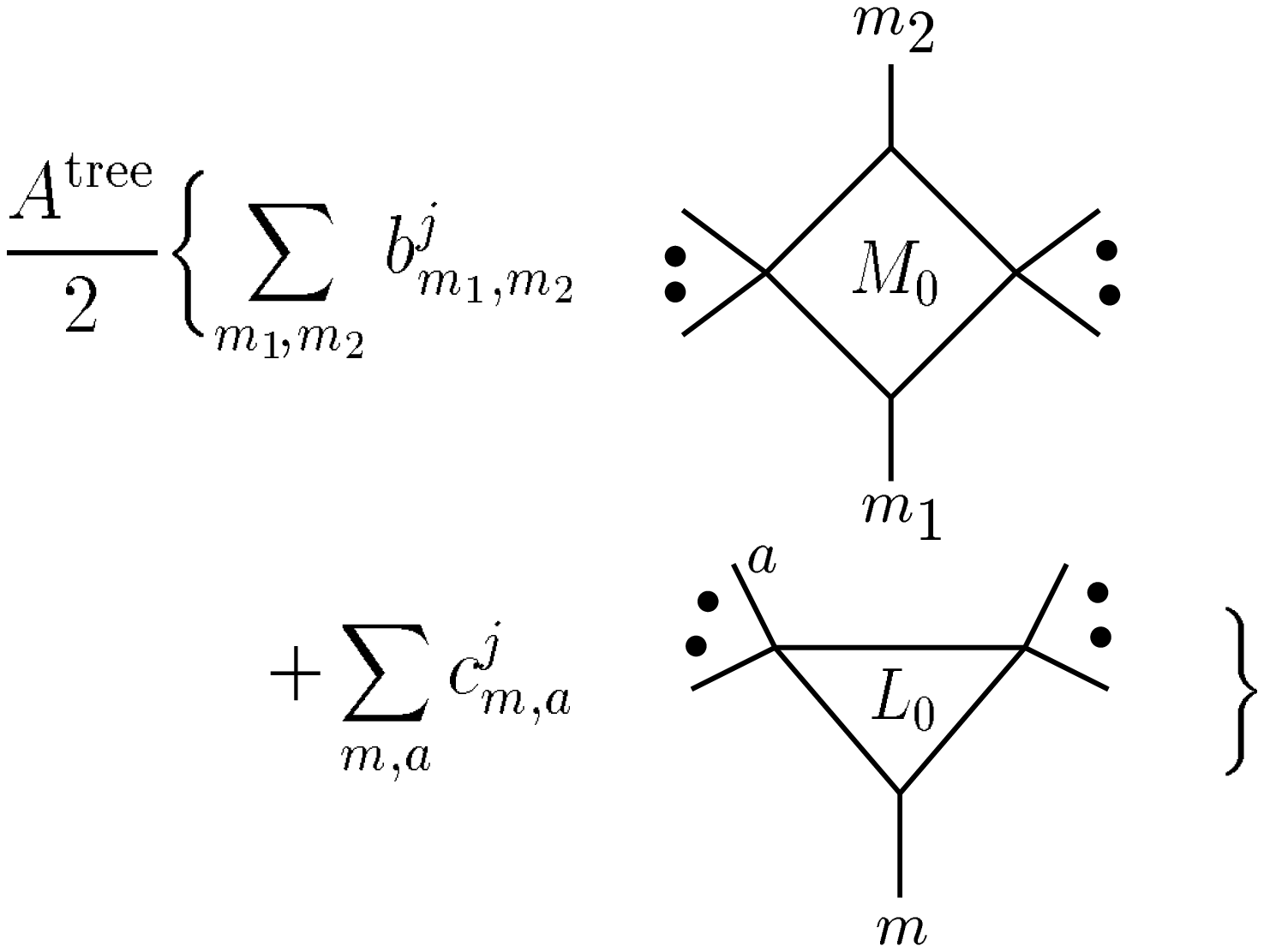}
\caption{The N=1 one-loop contributions to a generic MHV configuration $A(1^-,2^+,\cdots j^-,\cdots n^+)$. The
summation of $M_0$ terms runs over all (scaled and truncated) box functions where the negative helicity legs $1$ and $j$ lie in 
the massive legs. Explicitly $1 < m_1 < j < m_2$. The sum include the case where one massive leg is in fact null. }
	\label{non-adjacentMHV}
\end{figure}

 The six-point case is rather a degenerate form of the general case so it is useful to present here the explicit form
$$
\eqalign{
A(1^-& 2^+3^-4^+5^+6^+)
= { \cg \Atree \over 2 } \Bigl(
\cr
&
\hskip -1.0 truecm a^{13}_{62} \; {\cal F}^{1m} (s_{61},s_{12}, t_{345} ) 
+a^{13}_{24} \; {\cal F}^{1m} (s_{23},s_{34},t_{561} )
\cr
& +
a^{13}_{25} \; {\cal F}^{2me}(t_{234}, t_{345} , s_{34} , s_{61}  )  
\cr
&+%
c^{13}_{2,4}{\Lz( s_{34} /t_{234}  ) \over t_{234}   } +
c^{13}_{2,5}{\Lz( s_{61} / t_{612}  ) \over t_{612}     }
\cr
&+
c^{13}_{4,1}{\Lz( s_{23} / t_{234}  ) \over t_{234}  }+
c^{13}_{5,1}{\Lz( s_{61} / t_{561}  ) \over  t_{561}  }
\cr
&+
c^{13}_{5,2}{\Lz( s_{34} / t_{345}  ) \over  t_{345} }+
c^{13}_{6,2}{\Lz( s_{12} / t_{612}  ) \over  t_{612}  } )
\cr
&+
c^{13}_{2,6}{ \Kz( -s_{12}/\mu^2 ) \over s_{12} }  +
c^{13}_{6,1}{ \Kz( -s_{23}/\mu^2 ) \over s_{23} }  
\cr
&+
c^{13}_{4,2}{ \Kz( -s_{34}/\mu^2 ) \over s_{34} }  +
c^{13}_{2,3}{ \Kz( -s_{23}/\mu^2 ) \over s_{23} }  
\Bigr)
\cr}
\equn\label{MHV13sum}
$$
and
$$
\eqalign{
A(1^-& 2^+3^+4^-5^+6^+)
= { \cg \Atree \over 2 } \Bigl(
\cr
& \hskip -1.0 truecm
b^{14}_{62} \; 
{\cal F}^{1m} ( s_{61},s_{12},t_{345} )  
+b^{14}_{35} \; {\cal F}^{1m}(s_{34},s_{45},t_{612} ) 
\cr
 & \hskip 1.0 truecm +b^{14}_{63} \;{\cal F}^{2me}(t_{612},t_{123},s_{12},s_{45} )  
\cr
 & \hskip 1.0 truecm
 +b^{14}_{52} \;{\cal F}^{2me}(t_{561},t_{612},s_{61},s_{34} ) 
\cr
&+
c^{14}_{2,5}{\Lz( s_{61} / t_{612}   ) \over t_{612}   } +
c^{14}_{3,6}{\Lz( s_{12} / t_{123}   ) \over t_{123}     }
\cr
&+
c^{14}_{6,3}{\Lz( s_{45} / t_{456}  ) \over t_{456}  }+
c^{14}_{5,2}{\Lz( s_{34} / t_{345}  ) \over  t_{345}  }
\cr
&+
c^{14}_{6,2}{\Lz( s_{12} / t_{612}  ) \over  t_{612} }+
c^{14}_{5,1}{\Lz( s_{61} / t_{561}  ) \over  t_{612}  } )
\cr
&+
c^{14}_{2,4}{\Lz( s_{34} / t_{234}  ) \over  t_{234} }+
c^{14}_{3,5}{\Lz( s_{45} / t_{345}  ) \over  t_{345}  } )
\cr
&+
c^{14}_{2,6}{ \Kz( -s_{12}/\mu^2 ) \over s_{12} }  +
c^{14}_{6,1}{ \Kz( -s_{23}/\mu^2) \over s_{23} }  
\cr
&+
c^{14}_{4,3}{ \Kz( -s_{45}/\mu^2 ) \over s_{45} }  +
c^{14}_{2,4}{ \Kz( -s_{34}/\mu^2 ) \over s_{34} }  
\Bigr)
\cr}
\equn\label{MHV14sum}
$$

\section{$N=1$ amplitudes $A_6(--+-++)$, $A_6(-+-+-+)$ }

These amplitudes are not part of any known all-$n$ series.
Let us consider the case $(-+-+-+)$ first because it has the most symmetry being symmetrical under the operations
$$
\eqalign{
A & \longrightarrow A|_{i\longrightarrow i+2} 
\cr
\;\;A & \longrightarrow \bar{A}|_{i \longrightarrow i+1} 
\cr}
$$

 The amplitude contains all
six one-mass and all six two-mass-hard boxes.
$$
A(1^-2^+3^-4^+5^-6^+)_{\rm box}=
\sum_i a_{i}^{1m} {\cal F}_i^{1m}     
+
\sum_i
a_i^{2m} {\cal F}_i^{2m h}   
$$
We need only quote $a_1^{1m}$ and $a_2^{2m}$ since
$$
a_{i+2}=a_{i}|_{j\longrightarrow j+2} 
\;\;
a_{i+1}=\bar{a}_{i}|_{j\longrightarrow j+1} 
$$
with
$$
\eqalign{
a_4^{1m} =&
{ \BR{2}{5}^2 \BR15 \BR35 \over \spb1.3^2 \spa4.5 \spa5.6 \BR36 \BR14 K^2  } 
 \; K=K_{123}
\cr
a_5^{2m} =&
{ \BR{2}{5}^2  \BR35 \BR24
\over \spb1.2 \spa5.6 \BR36 \BR14 \BR34^2   }  \; K=K_{123}
\cr}
$$

The amplitude also has bubble integral functions, and, two, three-mass triangle functions
$$
\eqalign{
\sum_{i=1}^6 c_{i,2} I_2 (s_{ii+1}) +\sum_{i=1}^3 c_{i,3} I_2( t_{ii+1i+2} )
\cr
+b^{3m}_1  I_3^{3m} +b^{3m}_2  I_3^{3m} 
\cr}$$
The bubble functions are of two types depending upon whether they are bubbles in $s_{ii+1}$ or $t_{ii+1i+2}$
Again we need only specify one of each $c_i$ since
$$
\eqalign{
c_{i+2,a}&=c_{i,a}|_{j \longrightarrow j+2} \;\;
c_{i+1,a}=\bar{c}_{i,a}|_{j \longrightarrow j+1}
\;\;
\cr
b^{3m}_2&=\bar{b}^{3m}_1|_{j \longrightarrow j+1}
\cr}
$$

This amplitude (and $A_6(--+-++)$) were originally calculated in ref~\cite{Britto:2005ha} using the 
fermionic integration unitarity method. We choose to express these amplitude using functions which are 
explicitly rational in the (components) of the Weyl spinor~\cite{DPW}. 
 
To describe the bubble coefficients, it is useful to introduce
$$
H_1( a;b;K) \equiv   { \la b | K | a] \over  \la a | K | a] } =  { \la b | K | a] \over  a \cdot K  }
$$
$$
\eqalign{
G_1&(a;b,c;B; Q,K) \equiv
\cr
&
-{ [B|K(KQ-QK)|a\ra \la b | (KQ-QK)|c\ra 
\over 2 \la a | K Q | a \ra \Delta_3(K,Q) }
\cr
+&{ [B|K|a\ra( \spa{b}.{a} [a|K|c\ra +\spa{c}.{a} [a|K|d\ra )
\over 2 \la a | KQ | a \ra  [a|K|a\ra } 
\cr}
$$
where
$$
\Delta_3(K,Q) \equiv  
4(K\cdot Q)^2 -4K^2Q^2
$$
is the Gram determinant of the three mass triangle defined by having two legs with massive momenta $K$ and $Q$. 
Its appearance is a clear indication of the links between the bubble and triangle functions implied by the
absence of spurious singularities~\cite{spurious,BjerrumBohr:2007vu}. 

We also define extended versions $H_n$,$G_n$
$$
\eqalign{
&H_n ( a_1 \cdots a_n ; b_1 \cdots b_n ; K )=
\cr
& \hskip 0.5 truecm 
\sum_{i=1}^n    {  \prod_{j=2}^n \spa{b_j}.{a_i}  \over 
\prod_{j \neq i} \spa{a_j}.{a_i} }   H_1 (a_i,b_1;K)
\cr}
$$
$$
\eqalign{
&G_n ( a_1 \cdots a_n ; B, b_1 \cdots b_{n+1} ; K,Q)=
\cr
& \hskip 0.5 truecm 
\sum_{i=1}^n    {  \prod_{j=2}^n \spa{b_j}.{a_i}  \over 
\prod_{j \neq i} \spa{a_j}.{a_i} }   G_1 (a_i;B;b_1,b_{n+1};K,Q)
\cr}
$$

Using these functions we can express the bubble coefficients as
$$
\eqalign{
c_{1,3}=& 
-{  [2 | K|5\ra^2 \over \spa4.5\spa5.6\spb1.2\spb2.3 t_{123} }
\times 
\cr & H_4 (  4,6, K|3] , K|1] ; 5,5,K|2],K|2] , K )
\cr}
$$
where $K=K_{123}$,
and
$$
\eqalign{
&c_{1,2} = 
\cr
&{ [2|K|5\ra^2 \over \spb1.2\spa4.5\spa5.6 [3|K|6\ra t_{123} } 
\cr
&\hskip 0.5 truecm \times H_3(  2, K|3] , K'K|4\ra ;
1, K'K|5\ra ,  K'K|5\ra ; K' )
\cr 
&+{1 \over \spa1.2\spa3.4\spb5.6 }
\cr  &\hskip 0.5 truecm \times G_3( 2, Q|5] , K'K|4\ra ; 6 ;1,3 , X , X; Q'  ; K'  )
\cr
&+{[4|K_{345}|1\ra^2\over 
\spa1.2 \spb3.4\spb4.5[3|K_{345}|6\ra t_{345}}
 \cr &
\hskip 0.5 truecm
\times
H_3[ 2, 6,  K_{345}|5] ;  1 ,K_{345}|4], K_{345}|4] ;  K'];
\cr}
$$
where $K=K_{123},K'=K_{12},Q=K_{34}, Q'=K_{56}$ and 
$$
|X\ra = -|1\ra [6|K_{51}|3\ra -|2\ra\spb6.2\spa1.3
$$

This coefficient contains many spurious singularities.  Singularities $[a|K|a\ra$ generally cancel between
bubble functions  
$$
c I_2[ K^2]  +c' I_2[ (K+a)^2 ]  
\longrightarrow {\rm finite }
$$
The $\Delta^{-1}$ singularities vanish between the bubble coefficients and the three-mass triangle 
functions~\cite{spurious}

\def\Spa(#1,#2){\left\langle#1\,#2\right\rangle}
\def\Spb(#1,#2){\left[#1\,#2\right]}
\def\Spba(#1,\{#2,#3,#4\},#5){\left [#1|K_{#2#3#4}|#5\right \rangle}
\def\Spaa(#1,\{#2,#3\},\{#4,#5\},#6){\left\langle#1|K_{#2#3}K_{#4#5}|#6\right\rangle}
\def\Spab(#1,\{#2,#3,#4\},#5){\left [ #5|K_{#2#3#4}|#1\right \rangle}

The three-mass triangle coefficient is the most complicated function so far being, in the form given in 
ref.~\cite{BjerrumBohr:2007vu},
$$
\eqalign{ &b^{3m}_1 
= \cr &
 -\frac{\Spab(1,\{3,4,5\},4)\!
\Spab(1,\{3,4,5\},5)\! \Spab(6,\{3,4,5\},4)}{\Spab(2,\{3,4,5\},5)\!
\Spab(6,\{3,4,5\},3)\! \Spab(6,\{3,4,5\},5)\! t_{345}} 
\times \biggl(\!\Spa(3,5)\! \Spb(2,6)  
\cr
&\ + \frac{ \Spab(1,\{3,4,5\},4) (2 s_{12} s_{34}\! +\!(
s_{56}\! -\!s_{12}\!-\!s_{34}) t_{345})} {2 \Spa(1,2) \Spb(3,4)
\Spab(6,\{3,4,5\},5)} \biggr) 
\cr}
$$
$$
\eqalign{
& - \frac{ \Spab(2,\{5,6,1\},6)\! \Spab(3,\{5,6,1\},1)\!
\Spab(3,\{5,6,1\},6) }{\Spab(2,\{5,6,1\},1)\! \Spab(2,\{5,6,1\},5)\!
\Spab(4,\{5,6,1\},1) \!t_{561}}
\times \biggl( \Spa(5,1)\! \Spb(4,2) 
\cr
& +
\frac{ \Spab(3,\{5,6,1\},6) (2 s_{34} s_{56}\!+\!( s_{12}\! -\!s_{34}\!-\!s_{56}) t_{561})}{2 \Spa(3,4) \Spb(5,6) \Spab(2,\{5,6,1\},1)
}
\biggr)
\cr &
-
\frac{\Spab(4,\{1,2,3\},2)\! \Spab(5,\{1,2,3\},2)\! \Spab(5,\{1,2,3\},3)}{\Spab(4,\{1,2,3\},1)\! \Spab(4,\{1,2,3\},3)\! \Spab(6,\{1,2,3\},3) \!t_{123}}
\times \biggl( \!\Spa(1,3)\! \Spb(6,4)
\cr +
& \frac{\Spab(5,\{1,2,3\},2)
(2 s_{12} s_{56}\! -\!(s_{12}\!+\!s_{56}\!-\!s_{34}) t_{123})}{2 \Spa(5,6) \Spb(1,2)
\Spab(4,\{1,2,3\},3)  }
\biggr)
\cr &
\hskip -20pt -{ 1 \over \Delta_3(K_{12},K_{34} ) }\biggl(
\Spa(1,3) \Spa(3,5) \Spb(2,6) \Spb(3,4)\cr &+\Spa(1,3) \Spa(1,5) \Spb(1,2) \Spb(4,6)
+\Spa(1,5) \Spa(3,5) \Spb(2,4) \Spb(5,6)
\biggr)\times
\cr &
\hspace{-.6cm}\Biggl(\!\!
\frac{ 
[5|K_{34}|1\ra 
[4|K_{35}|6\ra
( t_{\!345}\!-\!t_{\!346\!} )}{
\Spab(2,\{\!3,4,5\!\},5)\!
\Spab(6,\{\!3,4,5\!\},3)\! 
\Spab(6,\{\!3,4,5\!\},5)}
\cr
\!+\! &
\frac{ \Spab(2,\{\!5,6,1\!\},6)
\!\Spab(3,\{\!5,6,1\!\},1)\! ( t_{\!561}\!-\!t_{\!562\!} )}
{\Spab(2,\{\!5,6,1\!\},1)\! \Spab(2,\{\!5,6,1\!\},5)\! \Spab(4,\{\!5,6,1\!\},1)}
\cr
&+ \!\frac{ \Spab(4,\{\!1,2,3\!\},2)\! \Spab(5,\{\!1,2,3\!\},3)\! (t_{\!123}\!-\!t_{\!124\!})}
{\Spab(4,\{\!1,2,3\!\},1)\! \Spab(4,\{\!1,2,3\!\},3)\! \Spab(6,\{\!1,2,3\!\},3)}
\cr 
-2   &
\frac{ 
[6|K_{34}|2\ra
[2|K_{56}|4\ra
[4|K_{12}|6\ra 
}{
\Spab(2,\{5,6,1\},5)
\Spab(4,\{4,5,6\},1)
\Spab(6,\{4,5,6\},3)}
\Biggr)
\cr 
+2   &
\frac{ 
[5|K_{34}|1\ra 
[1|K_{56}|3\ra 
[3|K_{12}|5\ra
}{
\Spab(2,\{5,6,1\},5)
\Spab(4,\{4,5,6\},1)
\Spab(6,\{4,5,6\},3)}
\Biggr)
\cr}
$$

The amplitude $A(--+-++)$ is a little more complicated since there is 
less symmetry amongst the coefficients.  The amplitude is symmetrical under the operation
$$
\eqalign{
A & \longrightarrow \bar{A}|_{123456\longrightarrow  654321} 
\cr}
$$

We can split up the amplitudes
$$
\eqalign{
A^{N=1}&(1^-2^-3^+4^-5^+6^+)
\cr
&= A|_{boxes} +A|_{three-mass-triangle}+A|_{bubbles} 
\cr}
$$

There are three two-mass hard boxes,and two one-mass boxes and the box
part of the amplitude is
$$
A|_{boxes}=a_1 \IIhard{4}+a_2\IIhard{6}+a_3\IIhard{2}+a_4\IIone{2}+a_5\IIone{3}
$$
where
$$
\eqalign{
a_1 
&= { [ 3 | K_{24} | 1 \rangle^2 
{  }
{ [ 3 | K_{234} | 5 \rangle }  \spa5.1
\over 
{ [ 4  | {K_{234}} | 5 \rangle^2 } 
{ [ 2  |{K_{234}} | 5 \rangle }  \spb2.3\spa5.6 \spa6.1 }
\cr
a_2
&=
{  [ 3 |{K}_{123} | 4  \rangle^2 
 \spb3.1 \spa6.4
\over [1  | {K}_{123} | 6 \rangle^2 \spb1.2\spb2.3\spa4.5\spa5.6  }
\cr
a_4
&= 
{  [ 3 | {K_{234}} |  1 \rangle^2
{ [ 2 | {K_{234}} | 1 \rangle }
\over
{ \spb2.4^2 \spa5.6 \spa6.1[ 2  |{K_{234}} | 5 \rangle K_{234}^2 }  
}
\cr
a_5 &=\bar{a}_4|_{123456\longrightarrow  654321}
\;\;
a_3=\bar{a}_1|_{123456\longrightarrow  654321}\cr}
$$
Note 
$$
\eqalign{
a_2= \bar{a}_2|_{123456\longrightarrow  654321}
\cr}
$$

The amplitude contains a single three mass triangle, that with massive legs $61$, $23$ and $45$.
The coefficient  of this is~\cite{BjerrumBohr:2007vu}
$$
\eqalign{
&b_{3m}^{[  \{2^- 3^+\}, \{4^- 5^+\},  \{6^+ 1^-\}  ]}=
\cr
&
{ \spb2.6   [2|K_{35}|4\ra  [6|K_{35}|4\ra     \over
\spb1.2    [2|K_{45}|3\ra  [2|K_{34}|5\ra   t_{345}                }
\times\!\Biggl(\!\spa1.2 \spb3.5+
\cr
&
{[6|K_{345}|4\ra ( 2 s_{61}s_{45} +(s_{23}-s_{61}-s_{45} )t_{345}  )\over
 2 \spb6.1\spa4.5  [2|K_{345}|3\ra }
\!\Biggr)\!
\cr
&
+{\spa5.1 [3|K_{234}|5\ra [3|K_{234}|1\ra\over \spa5.6  [4|K_{234}|5\ra  [2|K_{234}|5\ra t_{234}} 
\times \!\Biggl(\!\spa2.4  \spb6.5+
\cr
&
 { [3|K_{234}|1\ra (  2 s_{23}s_{61} +(s_{45}-s_{23}-s_{61})t_{234} )
\over
 2 \spb2.3\spa1.6 [4|K_{234}|5\ra   }
\!\Biggr)\!
\cr
&
+{   \spb1.3 \spa6.4  [3|K_{123}|4\ra\over
 \spb1.2 \spa5.6     [1|K_{123}|6\ra t_{123}              }
\times\!\Biggl(\!\spa1.2\spb6.5+
\cr
&
{[3|K_{123}|4\ra ( 2 s_{45}s_{23} +(s_{61}-s_{45}-s_{23})t_{123} )\over
  2 \spb2.3  \spa4.5 [1|K_{123}|6\ra }
\!\Biggr)\!
\cr}
$$
$$
\eqalign{ &
-{1 \over \Delta_3}\biggl(
\spa4.2\spa2.1\spb3.2\spb6.5 
\cr
&+
\spa4.1\spa2.1\spb6.1\spb3.5 +\spa4.2\spa4.1\spb4.5\spb3.6
\biggr)
\cr &
\times\Biggl(
2{      \spb2.6 \spa6.5 \spb3.6 \spa6.4 - \spa5.1 \spb1.2 \spa4.1 \spb1.3
   \over
        \spb1.2\spa5.6 [2|K_{61}|5\ra                                     }
\cr &
+{     \spb1.3  \spa4.6 (t_{123}-t_{623})
\over
       \spa5.6 \spb1.2  [1|K_{23}|6\ra  }
 +{   \spb2.6   [2|K_{345}|4\ra ( t_{345}-t_{245})
\over
     \spb1.2 [2|K_{34}|5\ra  [2|K_{45}|3\ra  }
\cr &+{     \spa5.1   [3|K_{234}|5\ra(t_{234}-t_{235})
\over
       \spa5.6 [2|K_{34}|5\ra  [4|K_{23}|5\ra}
\Biggr)\,,
\cr}
$$
The bubble part of this amplitude is
$$
\eqalign{
c_{2,2}  \Kz(-s_{23}/\mu^2 ) +c_{2,3}\Kz(-s_{34}/\mu^2 )  
\cr
+c_{2,4}\Kz(-s_{45}/\mu^2 )+c_{2,6}\Kz(-s_{61}/\mu^2  )
\cr
+c_{3,1} \Kz(-t_{123}/\mu^2  )+c_{3,2}\Kz(-t_{234}/\mu^2 )
\cr
+c_{3,3}\Kz(-t_{345}/\mu^2 ) 
\cr}
$$
From the symmetry of the amplitude we have
$$
c_{3,3}=\bar{c}_{3,2}|_{123456\longrightarrow 654321}\;\;\;
c_{2,4}=\bar{c}_{2,2}|_{123456\longrightarrow 654321}
$$
so have five bubble functions which we must define.
Firstly we have
$$
\eqalign{
c_{3,1}&=
{ - [3 | K_{123}|4\ra^2 \over \spa4.5\spa5.6\spb1.2\spb2.3 t_{123} }
\times 
\cr
&H_2 (  6, K_{123}|1] ; 4  ,K_{123}|3] , K_{123} )
\cr
c_{3,2}&=
{ - [3 | K_{234}|1\ra^2 \over \spa5.6\spa6.1\spb2.3\spb3.4 t_{234} }
\times 
\cr &
H_3 (  5, K_{234}|2] , K_{234}|4] ; 1, K_{234}|3],K_{234}|3] , K_{234} ) 
\cr}
$$
and
$$
\eqalign{
&c_{2,3}={ [3|K_{234}| 1\ra^2 \over
\spa5.6\spa6.1\spb3.4  [2|K_{234}|5\ra t_{234} }
\cr
&\hskip 1.0 truecm \times 
H_2( 3, K_{34}|2]; 4, K_{34}K_{56}|1]; K_{34} )
+
\cr
&{[6|K_{345}| 4\ra^2 \over
\spa3.4\spb6.1\spb1.2 [2| K{345}| 5\ra t_{345} }
\cr
&\hskip 1.0 truecm \times 
H_2[3, 5 ;  4, K_{12}| 6] ;  K_{34} ] 
\cr}
$$

$$
\eqalign{
&c_{2,2}=
-{ \spa2.4^2  \spb5.6^3 \over \spa2.3\spb6.1 [1| K_{56}| 4\ra t_{561} } \times
\cr
&
H_2[3, K_{61}|5] ; 2, 4;  K_{23} ]
\cr
&+
{ 1 \over \spa1.6\spa2.3\spb4.5 } \times
\cr
&
 G_4[3, K_{23}|4], K_{61}|5] , K_{23}K_{45}|6\ra ; 
\cr
&    \hskip 1.5truecm   5;  K_{23}|5], 2, 1, Y, Y ;  K_{45}; K_{23} ] 
\cr
&+
{ [3|K_{123}| 4\ra^2 \over  
\spa4.5\spa5.6\spb2.3 [1| K_{56}|4\ra t_{123}  }\times
\cr
&
\bar{H}_3[2, 1,   K_{45}|6\ra ;  3, K_{56}|4\ra , K_{56}|4\ra ; K_{23}     ] 
\cr}
$$
where
$$
|Y\ra = |2\ra [5|K_{24}|1\ra  +|3\ra \spa1.2\spb3.5
$$
and
$$
\eqalign{
&c_{2,6}=
\cr
&{ [6| K_{612}| 4\ra^2 \over   
\spa3.4\spa4.5\spb6.1[2|K_{612}| 5\ra t_{612}  }\times 
\cr
&
H_3[6, K_{612}|2] , K_{61}K_{612}|3\ra  ;
\cr & \hskip 1.0truecm 
 1,    K_{61}K_{612}|4\ra   ,K_{61}K_{612}|4\ra  ; K_{61} ] 
\cr
&
+{1 \over \spa1.6\spa2.3\spb4.5 }\times 
\cr
&
G_3[6, K_{23}|4], K_{61}K_{612}|3\ra;  5; 1, 2,  Z, Z; K_{45}, K_{61} ] 
\cr
&-{  [3| K_{612}| 1\ra^2  \over
\spa6.1\spb2.3\spb3.4 [2| K_{612}| 5\ra t_{612} }\times
\cr
&  H_3[6, 5, K_{234}|4];  1,  K_{234}|3 ],  K_{234}|3] ; K_{61}] 
\cr}
$$
where
$$
|Z\ra = -|1\ra [5|K_{41}|2\ra -|6\ra \spb5.6\spa1.2
$$

\section{$A^{[0]}_6(++++++)$}

This first of the scalar  amplitudes is relatively simple: it vanishes at tree level and consequently is purely rational at one-loop. 
It was originally deduced by examining collinear limits in ref.~\cite{Bern:1993qk} and consequently proven to be correct in 
\cite{Mahlon:1993si} using off-shell recursion. 
The general $n$-point form is 
$$
A_n(1^+\cdots n^+) =
{ i \cg \over 12}
{ E_n +O_n 
\over
\spa1.2\spa2.3\cdots \spa{n}.1 }
$$
where
$$
\eqalign{
E_n= - \sum_{1\leq i_1 < i_2 < i_3 < i_4 \leq n } 
\tr (\Slash{k}_{i_1}\Slash{k}_{i_2}\Slash{k}_{i_3}\Slash{k}_{i_4} )
\cr
O_n= - \sum_{1\leq i_1 < i_2 < i_3 < i_4 \leq n } 
\tr (\Slash{k}_{i_1}\Slash{k}_{i_2}\Slash{k}_{i_3}\Slash{k}_{i_4} \gamma_5  )
\cr}
$$
The two trace terms can obviously be combined as 
$2\tr_+(\Slash{k}_{i_1}\Slash{k}_{i_2}\Slash{k}_{i_3}\Slash{k}_{i_4} )$.  

For the six-point amplitude the above sum yields twelve terms.  
 
\section{$A^{[0]}_6(-+++++)$}
This amplitude also vanishes at tree level and consequently is purely rational at one-loop. 
It was first calculated in \cite{Mahlon:1993si} using off-shell recursion. We present the 
form~\cite{Bern:2005ji} which was obtained using  on-shell recursion
$$
\eqalign{
&A_{6}^{[0]}(1^-2^+3^+4^+5^+6^+) =  {i \cg  \over 6} \Biggl[ 
\cr  
& 
{   [6|K_{23}|1\ra^3 
   \over 
\spa1.2 \spa2.3 \spa4.5^2 [6|K_{12}|3\ra  t_{123}  }
\cr
& + {   [2|K_{34}|1\ra^3
    \over \spa3.4^2 \spa5.6 \spa6.1  [2|K_{34}|5\ra  t_{234}  }
\cr 
&+  { \spb2.6^3 \over \spb1.2 \spb6.1  t_{345} } \Biggl(
     { \spb2.3 \spb3.4 \over \spa4.5  [2 |K_{34}| 5\ra  }
    - { \spb4.5 \spb5.6 \over \spa3.4  [6|K_{12}|3\ra  }
\cr 
&     + { \spb3.5 \over \spa3.4 \spa4.5 } \Biggr)
- { {\spa1.3}^3 \spb2.3 \spa2.4
     \over {\spa2.3}^2 {\spa3.4}^2 \spa4.5 \spa5.6 \spa6.1 }
\cr 
&
+ { {\spa1.5}^3 \spa4.6 \spb5.6
     \over \spa1.2 \spa2.3 \spa3.4 {\spa4.5}^2 {\spa5.6}^2 }
\cr  
&
+
  {
\spa1.4^3 \spa3.5 [4|K_{23}|1\ra
    \over 
\spa1.2 \spa2.3 \spa3.4^2 \spa4.5^2 \spa5.6 \spa6.1 }
\Biggr].
\cr}
$$

\section{$A^{[0]}_6(---+++)$}

Surprisingly this NMHV amplitude is the simplest of the
``non-trivial'' scalar amplitudes. A general form is known and is
$$
\eqalign{
&A_n^{[0]}(1^-2^-3^-4^+5^+\cdots n^+)\; = 
\cr
&\; \frac{1}{3}\,A_{n}^{\,\NeqOne\ {\rm chiral}}
-{i\cg \over 3}
\sum_{r=4}^{n-1}\, \hat d_{n,r}\,
{    \Lt [ t_{3,r} / t_{2,r} ] \over t_{2,r}^3  }
\cr
&-{i\cg \over 3}
\sum_{r=4}^{n-2}\, \hat g_{n,r}\,
{    \Lt [ t_{2,r} / t_{2,r+1} ] \over t_{2,r+1}^3  }
 \cr &
-{i\cg \over 3}
\sum_{r=4}^{n-2}\ \hat h_{n,r}
{    \Lt [ t_{3,r} / t_{3,r+1} ] \over t_{3,r+1}^3  }
\cr
&  \hskip 2 cm +\cg \hat R_n\,, 
\cr}
$$
The functions 
$$
\eqalign{
\Lt(r) &= { \ln(r) -(r-r^{-1} )/2 \over (1-r)^3 }
\;\;
\cr
\Lone(r)&={ \ln(r) -(1-r) \over (1-r)^2 }
\cr}
$$ are non-singular as $r \longrightarrow 1$.  Using these functions includes
a part of the rational terms into the integral functions. 
$\hat R_n$ is the remaining rational terms CITE.

For the six-point the amplitude reduces to 
$$
\eqalign{& A_6^{[0]}(  1^-2^-3^-4^+5^+6^+) = {1\over 3 } A^{N=1} +\cr 
& - {i\over 2} \Biggl[ c_1 { \Lt [t_{345}/ s_{61} ] \over s_{61}^3 } +c_2{
\Lt [ t_{234} /s_{34} ] \over s_{34}^2}\cr 
& +c_3{ \Lt [ t_{234}/s_{61} ]
\over s_{61}^3 } +c_4{ \Lt [ t_{345} /s_{34} ] \over s_{34}^3 } 
\Biggr] +\cg \hat R_6 \cr}
$$
where the coefficients are
$$
\eqalign{ c_1 =& {[6|Kk_2K|3\ra[6|k_2|3\ra
[ 6 | ( k_2 K  -Kk_2) K|3\ra  \over
[2|K|5\ra  \spb6.1\spb1.2\spa3.4\spa4.5  }  
\cr}
$$
where $K=K_{345}$ and 
$$
\eqalign{ 
c_2 =& c_1|_{123456\longrightarrow 321654}
\;\;
c_4 = c_3|_{123456\longrightarrow 321654}
\cr
& \hskip 1.0 truecm 
c_3 = \bar{c}_1|_{123456\longrightarrow 456123}
\cr}
$$
and the rational term is
$$
\hat R_6 = X_6  +X_6|_{123456\longrightarrow 321654}
$$
where
$$
\eqalign{
X_6 &=
  {i\over6} { 1 \over \spb2.3 \spa5.6 \, [2|K_{34}|5\ra }
   \Biggl\{   - { {\spb4.6}^3 \spb2.5 \spa5.6 \over \spb1.2 \spb3.4 \spb6.1 }
 \cr
& - { {\spa1.3}^3 \spa2.5 \spb2.3 \over \spa3.4 \spa4.5 \spa6.1 }
-{ {\spa1.3}^2  ( 3  [4|K_2|1\ra +[4|K_3|1\ra  )
     \over \spa3.4 \spa6.1 }
\cr
 &+ {  [4|K_{23}|1\ra^2 \over \spb3.4 \spa6.1 }
          \biggl( { [4|K_2|1\ra  - [4|K_5|1\ra \over t_{234} }
          + { \spa1.3 \over \spa3.4 }
          - { \spb4.6 \over \spb6.1 } \biggl)
\cr
   & 
  + { {\spb4.6}^2 ( 3 [4|K_5|1\ra  + [4|K_6|1\ra )
     \over \spb3.4 \spb6.1 }
    \Biggr\}
\cr}
$$

\section{$A^{[0]}(--++++)$}

There exist general expressions for the MHV scalar amplitudes~\cite{Forde:2005hh,Berger:2006vq}. 
Within these the case of adjacent negative helicities simplifies
enormously and we have
$$
A^{[0]}(1^-2^-3^+4^+5^+6^+)={1 \over 3} A_6^{N=1} +{2\cg \over 9} \Atree_6 
+C_6+\cg \hat R_6
$$
where
$$
\eqalign{
C_6= -{\cg \Atree \over 3s_{12}^2 }
\Bigl(
& c_4  { \Lt (-s_{23}/-t_{234} ) \over t_{234}^3  }
\cr
+&c_5{ \Lt (-t_{561}/-s_{61} ) \over s_{61}^3  }
\Bigr) 
\cr}
$$
with
$$
\eqalign{
c_m=\tr[k_1k_2k_mq_{m,2}]\tr[k_1k_2q_{m,2}k_m]\cr
\times \tr[k_1k_2(q_{m,2}k_m-k_mq_{m,2})]
\cr}
$$
\def\sandp#1.#2.#3{%
\left\langle\smash{#1}{\vphantom1}^{-}\right|{#2}%
\left|\smash{#3}{\vphantom1}^{+}\right\rangle}
\def\sandpp#1.#2.#3{%
\left\langle\smash{#1}{\vphantom1}^{+}\right|{#2}%
\left|\smash{#3}{\vphantom1}^{+}\right\rangle}
\def\sandmm#1.#2.#3{%
\left\langle\smash{#1}{\vphantom1}^{-}\right|{#2}%
\left|\smash{#3}{\vphantom1}^{-}\right\rangle}
\def\spab#1.#2.#3{\sandmm#1.#2.#3}
\def\spba#1.#2.#3{\sandpp#1.#2.#3}
\def\spaa#1.#2.#3.#4{\sandmp#1.{#2#3}.#4}
\def\spbb#1.#2.#3.#4{\sandpm#1.{#2#3}.#4}
\def\spa#1.#2{\left\langle#1\,#2\right\rangle}
\def\spb#1.#2{\left[#1\,#2\right]}
\def\spash#1.#2{\vphantom{\hat K}\spa{\smash{#1}}.{\smash{#2}}}
\def\spbsh#1.#2{\vphantom{\hat K}\spb{\smash{#1}}.{\smash{#2}}}
\def\lor#1.#2{\left(#1\,#2\right)}
\def\sand#1.#2.#3{%
\left\langle\smash{#1}{\vphantom1}^{-}\right|{#2}%
\left|\smash{#3}{\vphantom1}^{-}\right\rangle}
\def\sandpp#1.#2.#3{%
\left\langle\smash{#1}{\vphantom1}^{+}\right|{#2}%
\left|\smash{#3}{\vphantom1}^{+}\right\rangle}
\def\sandpm#1.#2.#3{%
\left\langle\smash{#1}{\vphantom1}^{+}\right|{#2}%
\left|\smash{#3}{\vphantom1}^{-}\right\rangle}
\def\randmp#1.#2.#3.#4{%
\langle #1 |K_{#2} K_{#3}|#4\rangle}
\def\spab#1.#2.#3{[#3|K_{#2}|#1\ra}

\noindent
and where
$$
\eqalign{
& \hat R_6=
{1 \over 6} \Biggl\{ 
- 2 { \spa3.5 \spb3.5 
    \spab4.{12}.3 [4|K_{12}|6\ra  \spab5.{12}.6
  \over \spb1.2 \spa3.4^2 \spa4.5^2 \spb6.1
         \spab5.{34}.2 \spab6.{12}.3 } 
\cr}
$$
$$
 \eqalign{
&
- 2 { \spa3.5 \spb3.6 {\spab4.{12}.6}^2 
 \over \spb1.2 {\spa3.4}^2 {\spa4.5}^2 \spb6.1
        \spab5.{34}.2 } 
 \cr
&+ 2 {\spa1.2 \spa2.4 \spa3.5 
       {\spb3.5}^2 \spb5.6 \spab5.{(1+2)}.6
  \over {\spa3.4}^2 \spa4.5 \spb6.1
      \spab2.{16}.5 \spab5.{34}.2 \spab6.{12}.3 }
 \cr
&+ 2 { \spa1.2^2 \spb3.5^2
   \randmp5.{34}.{2}.1 + \randmp5.{3}.{5}.1 
   \over \spa3.4 \spa4.5 \spa6.1 
       \spab2.{16}.5 \spab5.{34}.2 \spab6.{12}.3 }
 \cr
&- { {\spa1.2}^3 \spa3.5 \spb4.6 \spb5.6
   \over \spa2.3 \spa3.4 \spa4.5 \spa5.6 
        \spab1.{23}.4 \spab3.{12}.6 }
\cr}
$$
$$
\eqalign{
&+ 2 { \spb3.6^3 
   \over \spb1.2 \spb2.3 \spa4.5^2 \spb6.1 }
\cr
&- { \spb5.6 {\spab5.{12}.6}^2
   ( 2 \randmp4.{35}.{12}.5 + \spa1.2 \spb1.2 \spa4.5 )
   \over \spb1.2 \spa3.4 {\spa4.5}^2 \spa5.6 \spb6.1
        \spab3.{12}.6 \spab5.{34}.2 } 
 \cr
&+ 2 { {\spa1.5}^2 {\spb3.4}^2 \spb5.6
   ( \spa1.6 \spb3.4 \spa4.5 - \spab1.{24}.3 \spa5.6 )
   \over \spb2.3 \spa4.5 {\spa5.6}^2 t_{234}
        \spab1.{23}.4 \spab5.{34}.2 }
 \cr
&- { \spa1.2 \spa1.5 \spb3.4 \spb5.6 
    \randmp1.{56}.{34}.5
   \over \spa3.4 \spa4.5 \spa5.6 t_{234}
    \spab1.{23}.4 \spab5.{34}.2 }
\cr
&+ 2 { \spa3.5 {\spab1.{24}.3}^3
   \over \spb2.3 \spa3.4 \spa4.5 \spa5.6 \spa6.1 t_{234}
       \spab5.{34}.2 }
 \cr}
$$
$$
 \eqalign{
&- { \spa1.2 \spab1.{24}.3 
       ( 2 \spab1.{24}.3 + \spab1.{4}.3 )
   \over \spb2.3 \spa3.4 \spa4.5 \spa5.6 \spa6.1 t_{234} }
 \cr
&+ 2 { {\spa1.2}^3 {\spb4.6}^2 \spab5.{46}.5
   \over \spa2.3 \spa4.5 \spa5.6 t_{123}
       \spab1.{23}.4 \spab3.{12}.6 }
\cr
&+ 2 { {\spa1.2}^3 {\spb3.5}^2 \spab4.{35}.4
   \over \spa3.4 \spa4.5 \spa6.1 t_{612}
       \spab2.{16}.5 \spab6.{12}.3 }
\cr
&- { {\spa1.2}^2
   \over \spa2.3 \spa3.4 \spa4.5 \spa5.6 \spa6.1 }
    \Biggl[ { \spab1.{4}.3 \over \spb2.3 }
          + { \spab2.{5}.6 \over \spb6.1 } \Biggr]
\Biggl\}
\cr}
$$

\begin{table}[ht]
\hrule
\def\tend{\cr \noalign{ \hrule}}
\halign{
       &  \vrule
       #   \strut        &\strut\hfil #\hfil\vrule
 \strut        &\strut\hfil #\hfil\vrule
 \strut        &\strut\hfil #\hfil\vrule
 \strut        &\strut\hfil #\hfil\vrule
 \strut        &\strut\hfil #\hfil\vrule
        \cr
height15pt  &  $\hbox{\small ++++++}$  & $\hbox{\small -- +++++}$        & $\hbox{\small -- -- ++++}$   & 
$\hbox{\small -- -- -- +++}$ &  $\hbox{\small -- + -- +++}$  
\tend
height15pt  & $1088$  & $1541$       &
$5210$   &  $1837$  & $20733$   
\tend 
}\caption[]{{\tt LeafCount} of the rational terms of some of the amplitudes. This is the {\tt LeafCount} of the
terms expressed as polynomial in $\lambda_a,\bar\lambda_{\dot a}$ without any attempt at simplification.
As such it is of comparative use only. The {\tt LeafCount} of the MHV tree amplitude, expressed equivalently is about 100 } 
\end{table}

\section{$A^{[0]}(-+-+++)$ and $A^{[0]}(-++-++)$}

The amplitude of course splits into cut-constructible and rational pieces. 
The cut constructible part of the amplitudes is just, for a MHV amplitude with negative helicities $i$ and $j$
$$
\eqalign{
& \Atree  \times \biggl( 
\sum (b_{ij}^{n_1n_2})^2  {\cal F}^{2me}_4 
+\sum \hat e^{ij}_{m,a} { \Lz  (s/s') \over s' }
\cr
&+\sum \hat f^{ij}_{m,a} { \Lone(s/s') \over (s')^2 }
+\sum g^{ij}_{m,a}{ \Lt(s/s') \over (s')^3 }\biggr)
+\hat R_6\cr}
$$
The box coefficients satisfy
$$
b^{[0]} =  { (b^{N=1})^2 \over b^{N=4} }
$$
a feature which is shared by the $A(--+-++)$ and $A(-+-+-+)$ amplitudes but which needs
extending for $n>6$ point amplitudes~\cite{Bidder:2004vx}. 
The coefficients of the ${\rm L}_i$ take the form,
$$
\eqalign{
\hat e^{ij}_{m,a} &= { 1 \over 3 } c^{ij}_{m,a} +I^{ij}_{m,a}
\cr
\hat f^{ij}_{m,a} &= { 1 \over 3 } S^{ij}_{m,m+a}
\cr
\hat g^{ij}_{m,a} &= { 1 \over 3 } { c^{ij}_{m,a} \over s_{ij}^2 }  \tr(k_ik_jk_mq_{m,a})\tr(k_ik_jq_{m,a}k_m)
\cr}
$$
where
$$
\eqalign{
S^{ij}_{m,a} &= {\tr_+(k_ik_j k_mk_{a+1})\tr_+(k_ik_j k_{a+1}k_m) \over s_{am+1}^2}
\cr
&-
{\tr_+(k_ik_j k_mk_{a})\tr_+(k_ik_j k_{a}k_m)\over s_{am}^2}
\cr
I^{ij}_{m,a} &= {\tr_+^2(k_ik_j k_mk_{a+1})\tr_+^2(k_ik_j k_{a+1}k_m) \over s_{am+1}^3}
\cr
&-
{\tr_+(k_ik_j k_mk_{a})\tr_+(k_ik_j k_{a}k_m)\over s_{am}^3}
\cr}
$$

For the six-point MHV amplitudes the summations of the ${\rm L}_i$ functions run over the same variables as in  
eq.~(\ref{MHV13sum}) and eq.~(\ref{MHV14sum}).  

The rational terms for these two amplitude we do not produce here. They are quite extensive.  The {\tt LeafCount} of these expression (as a naive rational function of $|k^{\pm}\ra$) is given in table~2. For comparison, the six-point MHV tree amplitude has a {\tt LeafCount} of
about 100. The {\tt LeafCount} of expressions is very sensitive to the way the functions are presented. For example
the general expression for the rational terms of $A(1^-2^-3^+\cdots n^+)$ when specialised to $n=6$ has a 
{\tt LeafCount} of 34642 as opposed to 5210 for the specialised form. The {\tt Mathematica} expressions of these  will be available to download.

\section{$A^{[0]}(--+-++)$ and $A^{[0]}(-+-+-+)$}

The cut-constructible parts of these amplitudes were calculated in ref.~\cite{Britto:2006sj}
using fermionic integration unitary methods. The rational terms were calculated using Feynman diagrams in ref.~\cite{Xiao:2006vt} with a particular notation. 
We shall not produce these amplitudes here.  The expressions are rather complicated and we have no 
innovative way to present these.  The results have been checked by comparison to the results for the 
six-gluon obtained using numerical methods~\cite{Ellis:2006ss}.

\section{Summary}
It is hoped that bringing together as much as practical, at this point, of the six-gluon amplitude is a useful exercise.
The amplitudes presented here will be available from {\it 
http://pyweb.swan.ac.uk/$\sim$dunbar/sixgluon.html}
in {\tt Mathematica} format. The amplitudes available there been tested against the numerical results of
ref.~\cite{Ellis:2006ss} and as such should be error free. It is intended to correct these as any typographic
(or other) problems arise. 

I would like to thank the organisers of {\it Loops and Legs in Quantum Field Theory, 2008}
both for their excellent organisation and for suggesting this contribution.   I am grateful to the authors of
refs.~[1-13] both for their original contributions and for many useful conversations over the years.


\begin{thebibliography}{99}


\bibitem{Mahlon:1993si}
  G.~Mahlon,
  Phys.\ Rev.\  D {\bf 49} (1994) 4438
  [arXiv:hep-ph/9312276].


\bibitem{Bern:1993qk} Z.~Bern, G.~Chalmers, L.~J.~Dixon and
  D.~A.~Kosower,  Phys.\ Rev.\ Lett.\ {\bf 72}
  (1994) 2134 [arXiv:hep-ph/9312333].  


\bibitem{BDDKa}
Z. Bern, L.J. Dixon, D.C. Dunbar and D.A. Kosower,
  Nucl.\ Phys.\  B {\bf 425} {1994} {217}, [arXiv:hep-ph/9403226], 


\bibitem{Bern:1994cg}
  Z.~Bern, L.~J.~Dixon, D.~C.~Dunbar and D.~A.~Kosower,
  Nucl.\ Phys.\  B {\bf 435} (1995) 59
  [arXiv:hep-ph/9409265].


\bibitem{Bidder:2004tx}
  S.~J.~Bidder, N.~E.~J.~Bjerrum-Bohr, L.~J.~Dixon and D.~C.~Dunbar,
  Phys.\ Lett.\ B {\bf 606}, 189 (2005)
[hep-th/0410296].


\bibitem{Bedford:2004nh}
  J.~Bedford, A.~Brandhuber, B.~J.~Spence and G.~Travaglini,
  Nucl.\ Phys.\  B {\bf 712} (2005) 59
  [arXiv:hep-th/0412108].


\bibitem{Britto:2005ha}
  R.~Britto, E.~Buchbinder, F.~Cachazo and B.~Feng,
  Phys.\ Rev.\  D {\bf 72} (2005) 065012
  [arXiv:hep-ph/0503132].

\bibitem{Bern:2005cq}
  Z.~Bern, L.~J.~Dixon and D.~A.~Kosower,
  Phys.\ Rev.\  D {\bf 73} (2006) 065013
  [arXiv:hep-ph/0507005].


\bibitem{Bern:2005hh}
  Z.~Bern, N.~E.~J.~Bjerrum-Bohr, D.~C.~Dunbar and H.~Ita,
  JHEP {\bf 0511} (2005) 027
  [arXiv:hep-ph/0507019].

\bibitem{Britto:2006sj}
  R.~Britto, B.~Feng and P.~Mastrolia,
  Phys.\ Rev.\  D {\bf 73} (2006) 105004
  [arXiv:hep-ph/0602178].



\bibitem{Berger:2006ci}
  C.~F.~Berger, Z.~Bern, L.~J.~Dixon, D.~Forde and D.~A.~Kosower,
  Phys.\ Rev.\  D {\bf 74} (2006) 036009
  [arXiv:hep-ph/0604195].


\bibitem{Berger:2006vq}
  C.~F.~Berger, Z.~Bern, L.~J.~Dixon, D.~Forde and D.~A.~Kosower,
  Phys.\ Rev.\  D {\bf 75} (2007) 016006
  [arXiv:hep-ph/0607014].

\bibitem{Xiao:2006vt}
  Z.~Xiao, G.~Yang and C.~J.~Zhu,
  Nucl.\ Phys.\  B {\bf 758}, 53 (2006)
  [arXiv:hep-ph/0607017].


\bibitem{ColorDecomposition}
J.E.\ Paton and H.M.\ Chan, Nucl.\ Phys.\ B {\bf 10} (1969) 516; 
F.A.\ Berends and W.T.\ Giele,  Nucl.\ Phys.\  B {\bf 294} (1987) 700;
M.\ Mangano,  Nucl.\ Phys.\  B {\bf 309} (1988) 461.


\bibitem{Colour}
  Z.~Bern and D.~A.~Kosower,
  Nucl.\ Phys.\  B {\bf 362} (1991) 389.



\bibitem{Xu:1986xb}
  Z.~Xu, D.~H.~Zhang and L.~Chang,
  Nucl.\ Phys.\  B {\bf 291} (1987) 392.
  
 
\bibitem{Passarino:1978jh}
  G.~Passarino and M.~J.~G.~Veltman,
  Nucl.\ Phys.\  B {\bf 160} (1979) 151.


\bibitem{Bidder:2005in}
  S.~J.~Bidder, D.~C.~Dunbar and W.~B.~Perkins,
  JHEP {\bf 0508} (2005) 055
  [arXiv:hep-th/0505249].


\bibitem{Bidder:2005ri}
  S.~J.~Bidder, N.~E.~J.~Bjerrum-Bohr, D.~C.~Dunbar and W.~B.~Perkins,
  Phys.\ Lett.\  B {\bf 612} (2005) 75
  [hep-th/0502028].


\bibitem{Kunszt:1994mc}
  Z.~Kunszt, A.~Signer and Z.~Trocsanyi,
  Nucl.\ Phys.\  B {\bf 420} (1994) 550
  [arXiv:hep-ph/9401294].


\bibitem{DPW}
  D.~C.~Dunbar, W.B. Perkins and E. Warrick,
in preparation
 


\bibitem{spurious}
  Z.~Bern, L.~J.~Dixon and D.~A.~Kosower,
  Nucl.\ Phys.\  B {\bf 513} (1998) 3
  [hep-ph/9708239];\\
  J.~M.~Campbell, E.~W.~N.~Glover and D.~J.~Miller,
  Nucl.\ Phys.\  B {\bf 498} (1997) 397
  [hep-ph/9612413].

\bibitem{BjerrumBohr:2007vu}
  N.~E.~J.~Bjerrum-Bohr, D.~C.~Dunbar and W.~B.~Perkins,
  JHEP {\bf 0804} (2008) 038
  [arXiv:0709.2086].


\bibitem{Bern:2005ji}
  Z.~Bern, L.~J.~Dixon and D.~A.~Kosower,
  Phys.\ Rev.\  D {\bf 72}, 125003 (2005)
  [arXiv:hep-ph/0505055].


\bibitem{Forde:2005hh}
  D.~Forde and D.~A.~Kosower,
  Phys.\ Rev.\  D {\bf 73} (2006) 061701
  [arXiv:hep-ph/0509358].

 

\bibitem{Bidder:2004vx}
  S.~J.~Bidder, N.~E.~J.~Bjerrum-Bohr, D.~C.~Dunbar and W.~B.~Perkins,
  Phys.\ Lett.\  B {\bf 608} (2005) 151
  [hep-th/0412023].


\bibitem{Ellis:2006ss}
  R.~K.~Ellis, W.~T.~Giele and G.~Zanderighi,
  JHEP {\bf 0605} (2006) 027
  [arXiv:hep-ph/0602185].


\end{thebibliography}
\end{document}